\documentclass[twocolumn,prb,,showpacs,amsmath]{revtex4}

\usepackage{amssymb}
\usepackage{mathrsfs}
\usepackage{graphicx}
\usepackage{float}
\usepackage{color}
\usepackage{ulem}

\begin{document}

\title{Majorana fermions in density modulated p-wave superconducting wires}
\author{Li-Jun Lang}
\affiliation{Beijing National Laboratory for Condensed Matter
Physics, Institute of Physics, Chinese Academy of Sciences,
Beijing 100190, China}

\author{Shu Chen}
\email{schen@aphy.iphy.ac.cn}
\affiliation{Beijing National Laboratory for Condensed
Matter Physics, Institute of Physics, Chinese Academy of Sciences, Beijing 100190,
China}

\date{\today}

\begin{abstract}
We study the p-wave superconducting wire with a periodically
modulated chemical potential and show that the Majorana edge
states are robust against the periodic modulation. We find that
the critical amplitude of modulated potential, at which the
Majorana edge fermions and topological phase disappear, strongly
depends on the phase shifts. For some specific values of the phase
shift, the critical amplitude tends to infinity. The existence of
Majorana edge fermions in the open chain can be characterized by a
topological $Z_2$ invariant of the bulk system, which can be
applied to determine the phase boundary between the topologically
trivial and nontrivial superconducting phases. We also demonstrate
the existence of the zero-energy peak in the spectral function of
the topological superconducting phase, which is only sensitive to
the open boundary condition but robust against the disorder.
\end{abstract}

\pacs{73.20.-r,73.63.Nm,74.78.Fk, 03.67.Lx}
\maketitle

\section{Introduction}
Searching for Majorana fermions (MFs) in
condensed matter systems has attracted intensive studies in past
years \cite{Wilczek,Beenakker}, due to the fundamental interest in
exploring the new type of particles fulfilling non-Abelian
statistics and the potential applications for the topological
quantum computing
\cite{Wilczek,Beenakker,Ivanov,Kitaev1,Kane,Nayak}. Among various
proposals for realizing the emergent MFs, the quantum wires with
p-wave pairings provide promising candidates for realizing
emergent MFs at the ends of wires. As shown in the original work
of Kitaev \cite{Kitaev1}, boundary MFs emerge in the
one-dimensional (1D) p-wave superconductor if the system is in a
topological phase. Moreover, quantum wires with a strong
spin-orbit coupling, or topologically insulating wires, subject to
a Zeeman magnetic field and in proximity of a superconductor, are
found to exhibit boundary MFs \cite{Lutchyn,Oreg}. Particularly,
experimental signatures of MFs in hybrid
superconductor-semiconductor nanowires have been reported very
recently \cite{Kouwenhoven}, which stimulates the study of
exploring MFs in 1D systems
\cite{Wimmer,Potter,Alicea,Shen,Sen,Loss,Zhang}. Schemes of
realizing Majorana chains have also been proposed in cold atomic
systems \cite{Jiang}, carbon nanotubes \cite{Klinovaja},
superconducting (SC) circuits \cite{You} and
quantum-dot-superconductor arrays \cite{Sau}.

As most of theoretical works focus on ideal homogeneous wires with
a uniform chemical potential, an important problem is the
stability of the MFs under the modulation of density and disorder.
In this work, we explore the nonuniform p-wave SC wires with a
periodic modulation of the chemical potential,  extending the
Kitaev's  p-wave SC model \cite{Kitaev1}, and study the fate of
the MFs under the density modulation and disorder. In general, a
periodically modulated potential is of benefit to the formation of
periodic density waves. If the modulation amplitude is large, one
may expect that the SC phase could be destroyed. As the robustness
of Majorana mode is protected by the SC gap, the MFs may be
unstable in the presence of density modulation and disorder. On
the other hand, recent work on the density modulated wires in the
absence of SC order parameter indicates the existence of
topologically protected edge states \cite{Lang,Kraus}. So far, it
is unclear whether the Majorana edge states is enhanced or
suppressed under the density modulation. The current work will
focus on this problem and show the robustness of Majorana edge
states against the density modulation and disorder. Particularly,
in some parameter regimes, we find that MFs always exist for the
arbitrarily strong modulation strength, which may shed light on
the design of quantum architectures producing robust MFs.

The rest of paper is organized as follows. We introduce the
density modulated p-wave superconductor model and derive the
Bogoliubov-de Gennes equations in Section \ref{model}, and
demonstrate the zero mode Majorana edge states, which strongly
depend on the phase shift, in spectra and wave functions in
Section \ref{Majorana}. In order to confirm the transition, we
define a $Z_2$ topological invariant in Section \ref{Z2}. Further
we use the definition of the invariant to derive the phase
boundaries in Section \ref{Pd}. The spectral function is
calculated in Section \ref{ssf} with a zero-energy peak detected
corresponding to the zero Majorana modes. Section \ref{summary}
gives a summary.

\section{Model of density modulated p-wave superconductors} \label{model}
We consider a typical lattice model of the 1D p-wave superconductor
with modulated chemical potentials, which is described by
\begin{eqnarray}
H = \sum_{i} [ (-t c_{i}^{\dag }c_{i+1}+ \Delta c_{i}c_{i+1}+
H.c.) -\mu _{i} c_{i}^{\dag }c_{i} ], \label{Ham}
\end{eqnarray}
where  $\hat{c}^\dagger_i$ ($\hat{c}_i$) is the creation
(annihilation) operator of fermions at the {\it i}-th site, $t$
the nearest-neighbor hopping amplitude, $\Delta$ the p-wave SC
order parameter and chosen real, and the modulated chemical
potential given by
\begin{equation}
\mu_{i}=V \cos(2\pi i \alpha+\delta)
\end{equation}
with $V$ being the strength, $\alpha=p/q$ a rational number ($p$
and $q$ are co-prime integers), and $\delta$ an arbitrary phase
shift. The modulated chemical potential can be generated by the
bichromatic optical lattice for cold atom systems
\cite{Jiang,Roati} or through the control of gates for quantum-dot
arrays and quantum wires \cite{Sau,Loss2}. In this work, we only
consider the case with $\alpha$ being a rational number. For the
incommensurate case with $\alpha$ being an irrational number, the
Hamiltonian (\ref{Ham}) with $\Delta=0$ reduces to the
Aubry-Andr\'{e} model \cite{Aubry}, for which the system undergoes
a delocalization to localization transition when $V>2t$.
Correspondingly, the SC system with nonzero $\Delta$ and
irrational $\alpha$ also undergoes a transition from a topological
phase to a topologically trivial localized phase when $V$ exceeds
a critical value \cite{Cai}.

In order to diagonalize the quadratic form Hamiltonian, we resort
to the Bogoliubove-de Gennes transformation \cite{Lieb,deGennes}
and define a set of new fermionic operators,
\begin{equation}
\eta _{n}^{\dag } = \frac{1}{2}\sum_{i=1}^{L}[(\phi _{n, i}+\psi
_{n,i})c_{i}^{\dag }+(\phi _{n,i}-\psi _{n,i})c_{i}],
\label{quasi}
\end{equation}
where $L$ is the number of lattice sites and $n=1, \cdots, L$. For
convenience, $\phi _{n, i}$ and $\psi _{n, i}$ are chosen to be
real due to the reality of all the coefficients in the Hamiltonian
(\ref{Ham}). In terms of the operators $\eta_n$ and
$\eta_{n}^{\dagger}$, the Hamiltonian (\ref{Ham}) is diagonalized
as $H=\sum_{n=1}^{L}\Lambda _{n}(\eta _{n}^{\dag }\eta
_{n}-\frac{1}{2})$,
where $\Lambda _{n}$ is the spectrum of the single
quasi-particles. From the diagonalization condition, $[\eta
_{n},H]=\Lambda _{n}\eta _{n}$, we can get the following coupled
equations:
\begin{eqnarray}
\Lambda _{n}\phi _{n, i} &=& ~~(\Delta -t)\psi _{n,i+1}-\mu
_{i}\psi _{n,i}
-(\Delta + t)\psi _{n,i-1},  \nonumber \\
\Lambda _{n}\psi _{n,i} &=&-(\Delta + t)\phi _{n,i+1}-\mu _{i}\phi
_{n,i} +(\Delta - t)\phi _{n,i-1}, \nonumber \\
\label{coupled}
\end{eqnarray}%
with $i=1,\ \cdots\ ,L$. From Eqs. (\ref{coupled}), one can prove
that if the equations have the solution of $(\phi _{n,i},\
\psi_{n,i})$ ($i=1,\ \cdots,\ L$) with a positive eigenvalue
$\Lambda _{n}>0$, $(\phi_{n, i},\ -\psi _{n,i})$ is also the
solution with the eigenvalue $ - \Lambda _{n}$, which implies $
\eta_n (\Lambda _{n})= \eta_n^{\dagger} (-\Lambda _{n})$.

\section{Boundary Majorana fermions} \label{Majorana}
The solution to Eqs. (\ref{coupled}) is related to the boundary
condition. The boundary Majorana states can only exist in the
system with open boundary conditions (OBC). To seek such a state,
we solve Eqs. (\ref{coupled}) under OBC of $\psi _{n,L+1}=\psi
_{n,0}=\phi _{n,L+1}=\phi _{n,0}=0$, and the Majorana edge states
correspond to the zero mode solution of $\Lambda_{n}=0$. For
$\Lambda_{n}=0$, $\phi_i$ and $\psi_i$ are decoupled and Eqs.
(\ref{coupled}) can be written in the transfer matrix form
\begin{eqnarray}
( \phi _{n,i+1}, \phi _{n,i} )^{T}  &=& A_{i} (
\phi _{n,i}, \phi _{n,i-1} )^{T} ,  \nonumber \\
( \psi _{n,i-1}, \psi _{n,i} )^{T}  &=& A_{i} ( \psi _{n,i}, \psi
_{n,i+1} )^{T},  \label{tfm}
\end{eqnarray}
with $A_{i}=\left(
\begin{array}{cc}
\frac{-\mu _{i}}{\Delta + t} & \frac{\Delta - t}{\Delta + t} \\
1 & 0%
\end{array}%
\right)$. To understand the boundary Majorana modes, we rewrite the
operators (\ref{quasi}) as
\begin{eqnarray}
\eta _{n}^{\dag } = \frac{1}{2}\sum_{i=1}^{L}[ \phi _{n, i}
\gamma_i^A  + i \psi _{n,i} \gamma_i^B ], \label{quasi2}
\end{eqnarray}
where $\gamma_i^A = c_{i}^{\dag }+c_{i}$ and $\gamma_i^B =
i(c_{i}-c_{i}^{\dag })$ are operators of two MFs, which fulfill
the relations $(\gamma_i^\alpha)^{\dagger} = \gamma_i^\alpha$ and
$\{\gamma_i^\alpha, \gamma_j^\beta\} = 2 \delta_{ij}
\delta_{\alpha \beta}$ with $\alpha$ and $\beta$ taking $A$ or
$B$. Coefficients of $\phi _{n,i}$ and $\psi _{n,i}$ in Eq.
(\ref{quasi}) are just the amplitudes of Majorana operators
$\gamma_i^A $ and $\gamma_i^B$, respectively. If there exists a
zero mode solution, according to Eqs. (\ref{tfm}), we will get a
decaying solution for one set of coefficients and a growing one
for the other. An example is the special case with $V=0$ and
$\Delta =t$, for which the zero mode solution is given by
$(\phi_{1},\phi_{2},\cdots,\phi_L)=(1,0,\cdots,0)$ and
$(\psi_{1},\cdots,\psi_{L-1},\psi_L)=(0,\cdots,0,1)$. In this
case, $\eta^{\dag } (\Lambda=0) = \frac{1}{2}[ \gamma_1^A + i
\gamma_L^B ]$, which means that the zero mode state is divided
into two separated MFs located at the left and right ends,
respectively.
\begin{figure}[tbp]
\includegraphics[width=1.0\linewidth]{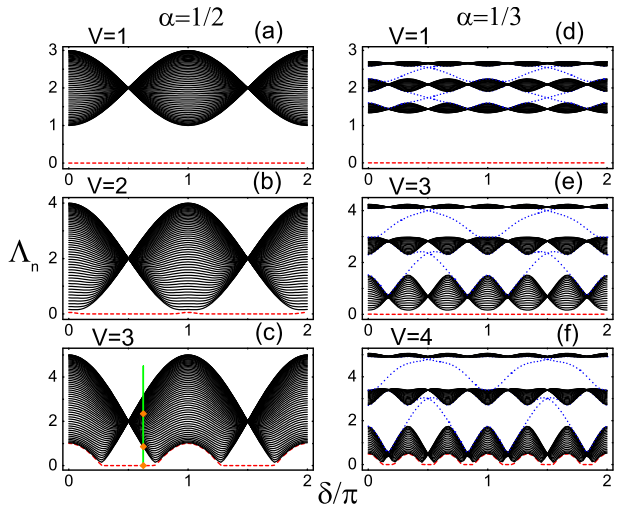}
\caption{The energy spectra, $\Lambda_n$, of 1D modulated p-wave
superconductors vs the phase shift, $\delta$, under OBC for
$\alpha=\frac{1}{2}$ ((a)-(c)) and $\frac{1}{3}$ ((d)-(f)). The
red dashed lines are the lowest excitation states, $\Lambda_1$,
and the blue dotted ones are the edge states in the higher
excitation gaps. These figures are calculated with parameters
$t=\Delta=1$ and $L=101$.} \label{spec}
\end{figure}

\begin{figure}[tbp]
\includegraphics[width=1.0\linewidth]{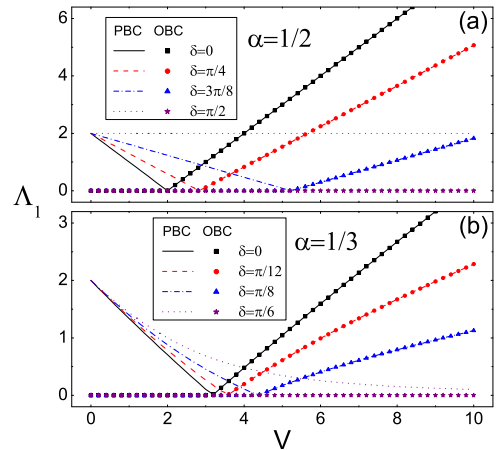}
\caption{The lowest excitation energies, $\Lambda_1$, as functions
of the modulation amplitude, $V$, at four typical phase shifts
(shown in Fig.) for the cases of $\alpha=\frac{1}{2}$ (a) and
$\frac{1}{3}$ (b) under PBC (lines) and OBC (symbols),
respectively. Here $t=\Delta=1$ and $L=510$.} \label{gap}
\end{figure}

For the commensurate potential with $\alpha =p/q$, $A_{i}$ is
$q$-periodic. One can judge whether the zero mode solution exists
by evaluating the two eigenvalues of the product matrix
$A=\prod_{i=1}^{q}A_{i}$ \cite{Sen}. If both of them are either
smaller or greater than unity, there is a zero mode solution with
MFs located at the ends, otherwise no zero mode solution is
available. Solving Eqs. (\ref{coupled}), we can get the whole
excitation spectrum for the single quasi-particles. To give
typical examples of $\alpha=1/2$ and $1/3$, we show the spectra
under OPC (Fig. \ref{spec}) and the lowest excitation energies,
$\Lambda _{1}$,  varying with $V$ under PBC and OBC (Fig.
\ref{gap}).  It is shown that under OBC, there are zero modes in
the bulk gap for the whole phase parameter space when $V$ is
small. As $V$ increases, the excitation gap shrinks; When it
exceeds a critical value, the gaps for some $\delta$'s close and
then reopen with zero modes vanishing, corresponding to a
transition from the topological phase to the topologically trivial
phase, and the regimes for the existence of the zero modes in the
phase parameter space become narrow. However, our numerical
results indicate that the zero mode Majorana solutions {\it
always} exist at some specific values of $\delta$, such as
$\delta=\pi/2$ for $\alpha=1/2$ and $\pi/6$ for $\alpha=1/3$, no
matter how strong the strength $V$ is (Fig.\ref{spec} and
Fig.\ref{gap}). To understand it, we observe that these $\delta$'s
have an exotic characteristic, that is, each of them makes the
value of $\cos(2\pi ip/q+\delta)$ zero for one of the sites, $i$,
in a supercell composed of $q$ sites, independent of the strength
of the modulated potential $V$. The fine tunability of the phase
shift $\delta$ effectively reduces the boundary potential and
makes it possible to exist Majorana boundary states even for very
large $V$. This reminds us that we have the exact Majorana edge
states for zero chemical potential in non-modulated systems
\cite{Kitaev1}. They have the same origin as the uniform case.
From the transfer matrix (\ref{tfm}), we can see the coefficients
are bounded to the ends of the open wire if there is a zero
modulated potential in a supercell. So these points are very
strong to have Majorana boundary states. Their exact expressions
can be derived from the boundary conditions in Section \ref{Pd}.
For the systems with $\alpha=1/2$ and $1/3$ shown in
Fig.\ref{spec}, the specific values of $\delta$ are
$\frac{(2m+1)\pi}{2}$ for $\alpha=1/2$ and $\frac{(2m+1)\pi}{6}$
for $\alpha=1/3$ with $m$ being an integer. At these specific
values $\delta_s$, the critical value of the phase transition,
$V_c(\Delta,\delta_s)$, which is a function of $\Delta$ and
$\delta$, tends to infinity. When deviated from these points, as
shown in Fig. \ref{gap}, there exists a finite critical value,
$V_c(\Delta,\delta)$, above which there appears a topologically
trivial phase without zero mode MFs.
\begin{figure}[tbp]
\includegraphics[width=1.0\linewidth]{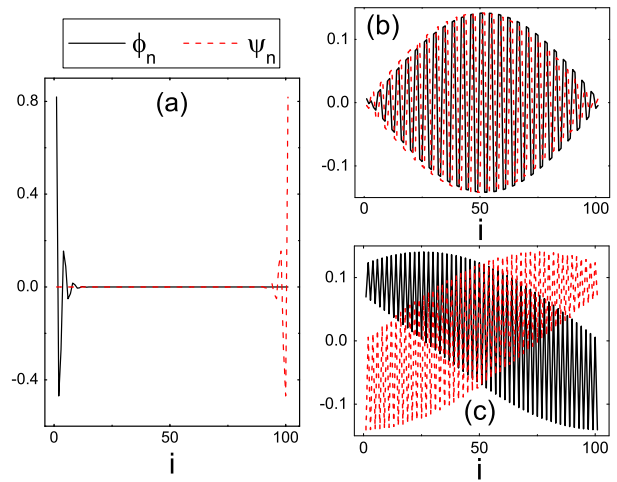}
\caption{The amplitudes, $\phi_n$ (black solid) and $\psi_n$ (red
dashed), for excitation states labelled by orange squares along
the green cut in Fig. \ref{spec}(c). From bottom to top, (a), (b),
and (c) represent the states of the zero mode, at the bottom and
at the center of the bulk band, respectively. They are calculated
with parameters $V=3,\ \alpha=\frac{1}{2},\ t=\Delta=1$,
$\delta=\frac{5\pi}{8}$ and $L=101$ under OBC.} \label{amp}
\end{figure}

Although we take $L=101$ to give a concrete example in Fig. \ref{spec}, we
note that the different length does not affect the existence of
the zero mode and the bulk energy shape, however it affects the
edge states in the higher excitation gaps \cite{Lang}. If we
select $L=102$, the only difference with the case of $L=101$ is
the left or right shifts of the edge states in the higher
excitation gaps for $\alpha=1/3$, the regimes with existence of
zero modes do not change.

In order to see clearly the differences between the zero mode
state and nonzero mode states, we display distributions of
$\phi_i$ and $\psi_i$ for the zero mode solution and solutions
with nonzero eigenvalues of quasi-particles located at the bottom
and center of the continuous band (orange squares along the green
cut in Fig. \ref{spec}(c)). As shown in Fig. \ref{amp}, for the
system with $\delta=5\pi/8$ and $V=3$, the amplitudes of the
Majorana operators for the zero mode solution are located at the
left and right ends, respectively. As $\phi_i$  ($\psi_i$) decays
very quickly away from the left (right) edge, there is no overlap
for the Majorana modes of $\gamma_i^A$ and $\gamma_i^B$. As a
comparison, distributions of $\phi_i$ and $\psi_i$ for nonzero
modes spread over the whole regime and the corresponding
quasiparticle operator can not be split into two separated
Majorana operators.

\section{$Z_2$ topological invariant} \label{Z2}
The presence or absence of zero mode MFs is determined by the
$Z_2$ topological class of the bulk superconductor. As no boundary
zero mode solution is available for the system with periodic
boundary conditions (PBC), we can choose a $Z_2$ topological
invariant (or `Majorana number') \cite{Kitaev1} to characterize
the topological nature of the bulk system. In order to define such
a topological invariant, we shall consider the system with PBC.
For the periodic system with $\alpha=p/q$, it is convenient to use
the Fourier transform,
$c_{i}=c_{s,l}=\sqrt{\frac{q}{L}}\sum_{k}c_{s,k}e^{ikql}$ with
$i=s+(l-1)q$, which transforms the Hamiltonian (\ref{Ham}) into
\begin{eqnarray} H_{k}=&&\sum_{s=1}^{q-1}(-tc_{s,k}^{\dag
}c_{s+1,k}+\Delta
c_{s,k}c_{s+1,-k})  \nonumber \\
&-&tc_{q,k}^{\dag }c_{1,k}e^{ikq}+\Delta c_{q,k}c_{1,-k}e^{-ikq}+H.c.  \nonumber \\
&-&\sum_{s=1}^{q}\mu _{s}(c_{s,k}^{\dag
}c_{s,k}-\frac{1}{2}),
\end{eqnarray}
where $s=1,\cdots,q$ is the index of inner sites in a supercell,
$l=1,\cdots,L/q$ the index of the $l$-th supercell, and $k$ the
momentum defined in the reduced Brillouin zone of $[0, 2\pi/q)$.
Then we define a set of new operators {\cite{Shen}}:
\begin{equation}
\gamma _{2s-1}(k)=c_{s,k}+c_{s,-k}^{\dag },\ \gamma
_{2s}(k)=(c_{s,k}-c_{s,-k}^{\dag })/i,  \label{newop}
\end{equation}
with the anticommutation relations $\{\gamma _{m}^{\dag
}(k),\gamma _{n}(k^{\prime })\}=2\delta _{mn}\delta _{kk^{\prime
}}$ and $\gamma _{m}^{\dag }(k)=\gamma _{m}(-k)$. $\gamma _{m}(0)$
and $\gamma _{m}(\pi /q)$ are just Majorana operators due to
$\gamma _{m}^{\dag }(0)=\gamma _{m}(0)$ and $\gamma _{m}^{\dag
}(\pi /q)=\gamma _{m}(-\pi /q)=\gamma _{m}(\pi /q)$. In the basis
of the new operators, we can transform the Hamiltonian into the
form:
\begin{equation}
H=\frac{i}{4}\sum_{k}\sum_{m,n}B_{m,n}(k)\gamma _{m}(-k)\gamma
_{n}(k),
\end{equation}
with $ B_{2s-1,2s}(k) = -B_{2s,2s-1}(k)=-\mu _{s}$ for $s=
1,\cdots ,q$, $ B_{2s-1,2s+2}(k) =-B_{2s+2,2s-1}(k)=\Delta - t$,
$B_{2s,2s+1}(k) =-B_{2s+1,2s}(k)=\Delta +t$ for $s = 1,\cdots
,q-1$, $ B_{1,2q}(k) =-B_{2q,1}^{\ast }(k)=-(\Delta +t)e^{-ikq}$,
and  $B_{2,2q-1}(k)=-B_{2q-1,2}^{\ast }(k)=-(\Delta -t)e^{-ikq}$.

\begin{figure}[tbp]
\includegraphics[width=1.0\linewidth]{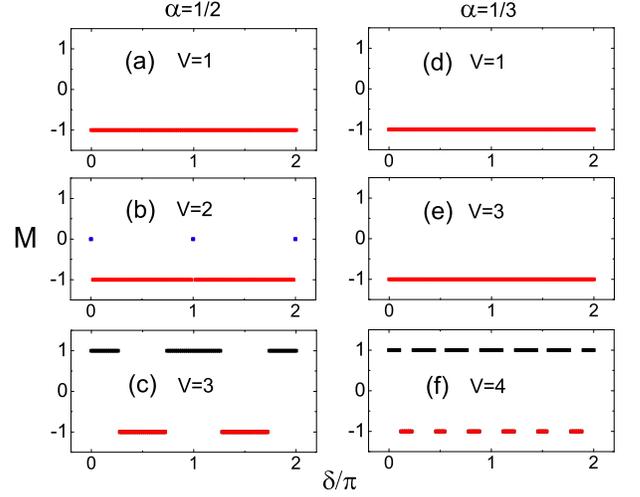}
\caption{$Z_2$ topological invariant, M, vs the phase shift,
$\delta$, for system with $t=\Delta=1$, $\alpha=\frac{1}{2}$
((a)-(c)) and $\frac{1}{3}$ ((d)-(f)) under PBC.} \label{fZ2}
\end{figure}

The parameters $B_{m,n}(k)$ form a $2q\times2q$ matrix $B(k)$, and
here only $B(0)$ and $B(\pi/q)$ are skew-symmetric. Following
Kitaev \cite{Kitaev1,Shen}, we can calculate the $Z_2$ topological
invariant defined as:
\begin{equation}
M=\text{sgn}[\text{Pf}(B(0))]\text{sgn}[\text{Pf}(B(\pi /q))],
\label{MaN}
\end{equation}%
where Pf$(X)=\frac{1}{2^NN!}\sum_P$sgn$(P)X_{P_1P_2}\cdots
X_{P_{2N-1}P_{2N}}$ is the Pfaffian of the skew-symmetic matrix
$X$ with $P$ standing for a permutation of $2N$ elements of $X$
and sgn$(P)$ the corresponding sign of the permutation. According
to the definition, generally we have $M =\pm 1$ with $M=1$
corresponding to a $Z_2$-topologically trivial phase and $M=-1$ to
a $Z_2$-topologically non-trivial phase. In Fig. \ref{fZ2}, we display the
$Z_2$ topological invariant versus the phase shift, $\delta$, for
systems with the same parameters as in Fig. \ref{spec}. Comparing
Fig. \ref{fZ2} with Fig. \ref{spec}, we see the exact correspondence between
the presence (absence) of zero mode MFs and $-1\ (+1)$ value of
the $Z_2$ topological invariant. Especially, $M$ takes the value
of $0$ (blue square dots in Fig. \ref{fZ2}(b)) at $V=2$, which is just the
critical point $V_c$ of the phase transition from a
$Z_2$-topologically non-trivial phase to a topologically trivial
phase for the corresponding $\delta$. In principle, we can always
determine $V_c(\Delta,\delta)$ through the condition of $M=0$ and
give the whole phase diagram.

\section{Phase diagram} \label{Pd}
\begin{figure}[tbp]
\includegraphics[width=1.0\linewidth]{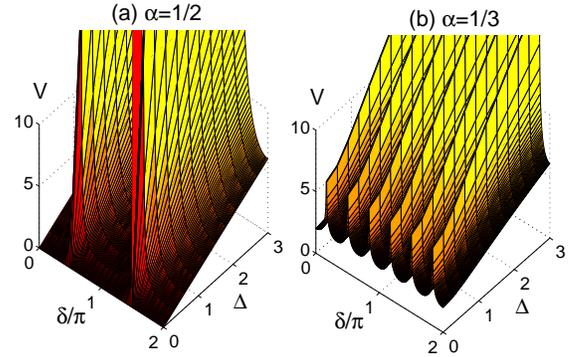}
\caption{Phase diagrams expanded by $V$, $\Delta$, and $\delta$
for the cases of (a) $\alpha=\frac{1}{2}$ and (b) $\alpha=\frac{1}{3}$. Here we
set $t=1$. The curved surfaces are the phase boundaries above
which is $Z_2$-topologically trivial while below which is
$Z_2$-topologically non-trivial.} \label{pd}
\end{figure}

Without loss of generality, we choose
$\Delta,\ V\geq0$ in the following discussion. From the definition
of the matrix $B(k)$, we know that for $\alpha=1/2$,
Pf$[B(0)]=-V^{2}\cos^{2}\delta -4t^{2}$ and Pf$[B(\pi
/2)]=-V^{2}\cos^{2}\delta +4\Delta ^{2}$. Here due to
Pf$[B(0)]<0$, we just need to set Pf$[B(\pi /2)]=0$ to get the
phase boundary condition, that is, $V^{2}\cos^{2}\delta =4\Delta
^{2}$, or equivalently $V|\cos\delta |=2\Delta $ (Fig.
\ref{pd}(a)). Likewise, for $\alpha=1/3$,
Pf$[B(0)]=-\frac{V^{3}}{4}\cos 3\delta -2t(t^{2}+3\Delta ^{2})$
and Pf$[B(\pi /3)]=-\frac{V^{3}}{4}\cos 3\delta +2t(t^{2}+3\Delta
^{2})$. The phase boundary condition can be also got easily as
$V^{3}|\cos 3\delta |=8t(t^{2}+3\Delta ^{2})$ (Fig. \ref{pd}(b)).
From Fig. \ref{pd},
we see clearly again that there exist some specific points
$\delta_s$, at which the system is always $Z_2$-topologically
non-trivial for arbitrary $V$. On the other hand, the critical
value, $V_c(\Delta,\delta)$, increases with the increase of the SC
pairing amplitude $\Delta$.

For the general case of $\alpha =p/q$, we can infer the phase
boundary condition by analyzing the expression of Pfaffian. The
non-permutated term, $B_{12}B_{34}\cdots B_{2q-1,2q}$, gives
$(-1)^q\prod_{s=1}^{q}\mu _{s}$. Due to the sparsity of the matrix
$B$, we see that there are only three kinds of permutations which
contribute non-zero terms: 1. $B_{2s-1,2s}B_{2s+1,2s+2}\rightarrow
B_{2s-1,2s+2}B_{2s,2s+1}$ or $B_{1,2}B_{2q-1,2q}\rightarrow
B_{1,2q}B_{2,2q-1}$, which makes a replacement of $\mu _{s}\mu
_{s+1}$ or $\mu _{1}\mu _{q}$ in the non-permutated term by
$(\Delta^{2}-t^{2})$ or $(\Delta ^{2}-t^{2})e^{-2ikq}$,
respectively; 2. $B_{2s-1,2s}\rightarrow B_{2s,2s+1}$
$(s=1,\cdots,q-1)$ and $B_{2q-1,2q}\rightarrow B_{1,2q}$, which
gives $-(t+\Delta )^{q}e^{-ikq}$; 3. $B_{2s-1,2s}\rightarrow
B_{2s-1,2s+2}$ $(s=1,\cdots,q-1)$ and $B_{2q-1,2q}\rightarrow
B_{2,2q-1}$, which gives $-(t-\Delta )^{q}e^{-ikq}$. These
permutations are independent of each other. Our numerical study
shows that when $k=0$ and $\pi /q$, the sum of all terms generated
by the first kind of permutations is zero except the
full-permutated ones (if there is) without any $-\mu _{s}$ left.
So the first kind contributes a term of $0\ $ for odd $q$'s  or
$(\Delta ^{2}-t^{2})^{q/2}+(\Delta ^{2}-t ^{2})^{q/2}e^{-2ikq}\ $
for even $q$'s. Thus we have Pf$[B(0)]=-\prod_{s=1}^{q}\mu
_{s}-[(t+\Delta )^{q} + (t-\Delta )^{q}]$  for odd $q$'s or
$\prod_{s=1}^{q}\mu _{s}-[(t+\Delta )^{q/2}-(-1)^{q/2}(t-\Delta
)^{q/2}]^{2}$ for even $q$'s, and Pf$[B(\pi
/q)]=-\prod_{s=1}^{q}\mu _{s}+[(t+\Delta )^{q}+(t-\Delta )^{q}]$
for odd $q$'s or $\prod_{s=1}^{q}\mu _{s}+[(t+\Delta
)^{q/2}+(-1)^{q/2}(t-\Delta )^{q/2}]^{2}$ for even $q$'s.
Therefore, we can get the general formula for the boundaries: For
$q$ is odd,
\begin{equation}
\left\vert \prod_{s=1}^{q}\mu _{s}\right\vert =(t+\Delta )^{q}+(t-\Delta
)^{q},
\end{equation}
and for $q$ is even,%
\begin{equation}
-\prod_{s=1}^{q}\mu _{s}=(t+\Delta )^{q}+(t-\Delta )^{q}+2(\Delta
^{2}-t^{2})^{q/2}
\end{equation}
if $\prod_{s=1}^{q}\mu _{s}<0$, or
\begin{equation}
\prod_{s=1}^{q}\mu _{s}=(t+\Delta )^{q}+(t-\Delta )^{q}-2(\Delta
^{2}-t^{2})^{q/2}
\end{equation}
if $\prod_{s=1}^{q}\mu _{s}>0$. The special cases of $\alpha =1/2$
and $1/3$ can be derived from these general formula.
\begin{figure}[h]
\includegraphics[width=1.0\linewidth]{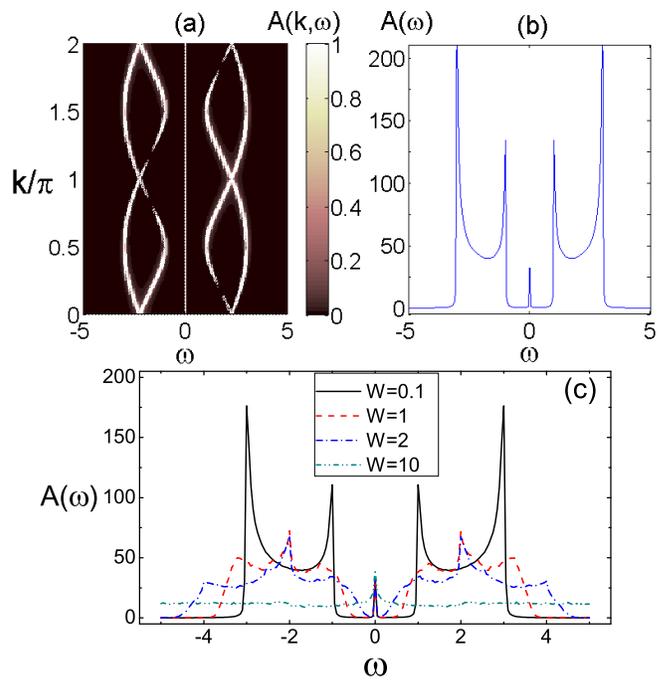}
\caption{The spectral functions and the disorder effects. Here, as
example, we choose $\protect\alpha =\frac{1}{2},\ \protect\delta =0$ to
plot (a) the momentum-resolved spectral function,
$A(k,\protect\omega )$, and (b) the corresponding total spectral
function, $A(\protect\omega )$, without
disorder, i.e. $W=0$. (c) shows that the total spectral function, $A(\protect%
\omega )$, varies with the strength of disorder.
The other parameters are $L=250,t=\Delta =V=1$ under OBC. We have
done $500$ realizations to calculate the disorder effects.}
\label{sf}
\end{figure}

The above results show that the phase boundary formula is
different for $q$ being even or odd. We note that the even-odd
effect comes from the cosine form of the modulated potential, the
values of which have a mirror symmetry with respect to zero value
if we choose even number of sites in one period. That does not
affect the fact of the existence of these special $\delta_s$ where
the critical amplitude $V_c$ tends to infinity. Only the number of
these points are different, i.e., $q$ for even cases and $2q$ for
odd cases.

\section{Spectral function} \label{ssf}
As the Majorana edge states
correspond to the zero mode solution protected by the presence of
an energy gap, one would expect a zero-energy peak appearing in
the corresponding spectral function. The
momentum-resolved spectral function with momentum $k$ and energy $\omega $ ($%
\hbar =1$) is defined as $A(k,\omega )=-\frac{1}{\pi }\text{Im}%
G_{r}(k,\omega )$, where $G_{r}(k,\omega )=\int_{-\infty }^{\infty
}G_{r}(k,t)e^{i\omega t-0^{+}t}dt$ and $G_{r}(k,t)=-i\theta
(t)\left\langle G\right\vert \{c_{k}(t),c_{k}^{\dag
}(0)\}\left\vert G\right\rangle $ is the single particle
(retarded) Green function with $\left\vert G\right\rangle $ being
the ground state of the system, and
$c_{k}(t)=e^{iHt}c_{k}e^{-iHt}$ is the fermionic annihilation
operator in the moment space with
$c_{k}=\frac{1}{\sqrt{L}}\sum_{j}c_{j}e^{-ikj}$. And the
corresponding total spectral function is
$A(\omega)=\sum_kA(k,\omega)$. As shown in Fig. \ref{sf}(a) and
(b), an obvious zero-energy peak is observed for the system with
OBC. On the contrary, we find no zero-energy peak in the system
with PBC.

The Majorana edge states are expected to be immune to local
perturbations. To explore the effect of disorder, here we add a
random on-site potential, $H_{\text{D}}= \sum_i W_ic_i^{\dagger}
c_i$, to the Hamiltonian (\ref{Ham}), which usually leads to
Anderson localizations, where $W_i$ is uniformly distributed in
the range of $[-W/2,W/2]$ with $W$ being the disorder strength.
Fig. \ref{sf}(c) shows that energy gaps between the zero mode and
the excitation modes are smeared when the disorder strength is
comparable to the excitation gap. Even in the presence of strong
disorder, for example, $W=10$, the peak still exists, although its
relative height becomes shorter as well. Meanwhile, the bulk
energy bands are broadened by the disorder, and other peaks are
not stable under disorder as the zero peak. Finally, they will be
smeared by the disorder with the increase in disorder strength.
The robustness of the zero peak against local perturbations
provides the possibility to detect it in experiment even affected
by environment disorder.

\section{Summary} \label{summary}
In summary, we find that in the density modulated p-wave SC wires,
by tuning the phase shift, the zero mode Majorana edge states
emerge in some intervals, and some of them are strong enough,
independent of the strength of the periodic density modulation.
The appearance of the MFs demonstrate the $Z_2$ topological
nature. After calculating the $Z_2$ topological invariant, we get
the phase diagram for the transition from topologically nontrivial
phases to trivial phases. By this model, we supply a good platform
to have possible schemes of searching for MFs. At last, we also
give an evidence in the spectral function where a zero-energy peak
appears under OBC, even subject to disorder. The spectral function
is possible to be experimentally detected by the photoemission
spectroscopy.

\begin{acknowledgments}
This work has been supported by National Program for Basic
Research of MOST, NSF of China under Grants No.11121063 and
No.11174360, and 973 grant.
\end{acknowledgments}


\end{document}